\documentclass[11pt]{article}
\usepackage{epsfig}
\usepackage{amssymb}
\newcommand{\bA}{{\mathbb A}}
\newcommand{\bB}{{\mathbb B}}
\newcommand{\bC}{{\mathbb C}}

\newcommand{\bN}{{\mathbb N}}

\newcommand{\bP}{{\mathbb P}}

\newcommand{\bR}{{\mathbb R}}

\newcommand{\bX}{{\mathbb X}}

\newcommand{\al}{\alpha}
\newcommand{\be}{\beta}
\newcommand{\ga}{\gamma}
\newcommand{\de}{\delta}

\newcommand{\et}{\eta}
\newcommand{\ze}{\zeta}
\renewcommand{\th}{\theta} % stupid LateX 2e !!!!!

\newcommand{\la}{\lambda}

\newcommand{\si}{\sigma}

\newcommand{\ph}{\phi}

\newcommand{\ps}{\psi}
\newcommand{\om}{\omega}
\newcommand{\vphi}{\varphi}

\newcommand{\veps}{\varepsilon}
\newcommand{\Ga}{\Gamma}
\newcommand{\De}{\Delta}

\newcommand{\Ps}{\Psi}

\newcommand{\bGa}{{{\rm I}\kern-.16em \Gamma}}
\newcommand{\etq}{{\bar\eta}}

\newcommand{\psq}{{\bar\psi}}

\newcommand{\Psq}{{\bar\Ps}}
\newcommand{\cA}{{\cal A}}

\newcommand{\cG}{{\cal G}}
\newcommand{\cH}{{\cal H}}

\newcommand{\cK}{{\cal K}}
\newcommand{\cL}{{\cal L}}
\newcommand{\cM}{{\cal M}}

\newcommand{\cR}{{\cal R}}
\newcommand{\cS}{{\cal S}}
\newcommand{\cT}{{\cal T}}
\newcommand{\cU}{{\cal U}}

\newcommand{\del}{\partial}
\newcommand{\nato}[1]{\{ 1,\ldots,#1 \} }

\newcommand{\abs}[1]{{\left\vert #1 \right\vert}}
\newcommand{\norm}[1]{{\left\Vert #1 \right\Vert}}

\newcommand{\Ref}[1]{$(\ref{#1})$}
\newcommand{\sfrac}[2]{{\textstyle \frac{#1}{#2}}}

\newcommand{\dotcup}{\mathop{\mathop{\cup}\limits^\cdot}}

\newcommand{\ili}{\int\limits}
\newcommand{\sli}{\sum\limits}
\newcommand{\pli}{\prod\limits}
\newcommand{\lli}{\lim\limits}
\newif\ifintrmk

\newcommand{\noin}{\noindent}
\newcommand{\nonu}{\nonumber}

\newcommand{\tst}{\textstyle}

\newcommand{\E}{{\rm e}}
\newtheorem{theorem}{Theorem}   

\newtheorem{lemma}{Lemma}
\newenvironment{proof}{\par\noin {\it Proof:} \hspace{7pt}}%
{\hfill\hbox{\vrule width 7pt depth 0pt height 7pt}
\par\vspace{10pt}}
\newcommand{\dd}{{\rm d}}
\newcommand{\ul}[1]{\underline{#1}}

\newcommand{\CC}{{\Ga}}
\newcommand{\DDe}{{\De\mkern-14mu\De}} %for thicker bf, take 13.9
\newcommand{\qq}{{\bf q}}
\newcommand{\sss}{{\bf s}}
\newcommand{\connpart}[1]{{#1}_{\mathrm c}}
\newcommand{\Xspace}{\bX}
\newcommand{\Bspace}{\bB}
\newcommand{\bb}{\xi}
\newcommand{\barm}{{\overline{m}}}

\newcommand{\truncated}[1]{#1} %%%%%{{#1}^T}

\renewcommand{\etq}{{\overline{\et}}}
\renewcommand{\Psq}{{\overline{\Ps}}}
\newcommand{\AbsC}{{\bf C}}
\newcommand{\pseudo}{d}
\newcommand{\tomC}{{\tilde\om_C}}
\newcommand{\ellC}{{\ell_C}}

\newif\ifpenrose
\penrosetrue

\begin{document}

%\rightline{MS--TPI--99--08}
\rightline{August 1999}

\vspace{3cm}
\begin{center}
{\LARGE\bf
Positivity and convergence \\
in fermionic quantum field theory \\[20mm]} 

{\large Manfred Salmhofer$\;^a$ and Christian Wieczerkowski$\;^b$} \\[10mm]
\small
$^a$ Mathematik, ETH--Zentrum, 8092 Z\" urich, Switzerland \\
\small
$^b$
Institut f\" ur Theoretische Physik I, Universit\" at M\" unster,
\\
\small
D--48149 M\" unster, Germany\\[10mm]
\end{center}
\normalsize

\begin{abstract}
\noindent
We derive norm bounds that imply the convergence of perturbation
theory in fermionic quantum field theory if the propagator is summable
and has a finite Gram constant. 
These bounds are sufficient for an application in renormalization group
studies. Our proof is conceptually simple and technically
elementary; it clarifies how the applicability of Gram bounds with
uniform constants is related to positivity properties of matrices
associated to the procedure of taking connected parts of Gaussian
convolutions. This positivity is preserved in the decouplings that
also preserve stability in the case of two--body interactions.
\end{abstract}
\thispagestyle{empty}

\newpage
\setcounter{page}{1}

\section{Introduction}

In fermionic field theories with an infrared and an ultraviolet
cutoff, perturbation theory converges. Perturbation theory in
bosonic theories always diverges. When representing the
perturbation series in terms of Feynman graphs, this is often
stated in the way that, although there are as many Feynman graphs
contributing to the fermionic perturbation expansion as to the
bosonic one, there are sign cancellations due to the fermionic
antisymmetry that lead to convergence. This explanation is
correct, but it is another matter to make the cancellations
explicit in a way that one can use them to remove the cutoffs in
nonperturbative constructions of fermionic models. This was done
for the Gross--Neveu model in 
\cite{Gawedzki/Kupiainen:1985} and
\cite{Feldman/Magnen/Rivasseau/Seneor:1985}.
Recently, there have been various results
that make these arguments more explicit
\cite{Abdesselam/Rivasseau:1995,Abdesselam/Rivasseau:1997} and
further simplify them, in particular by avoiding
cluster--expansion type arguments
\cite{Brydges/Wright:1988,Feldman/Knoerrer/Trubowitz:1998,Disertori/Rivasseau:1998}.
Our results go in the same direction, but we believe that they
provide an additional structural understanding of how the sign
cancellations occur and further reduce the technicalities in the
proofs. In our proof, we do not need to expand in Feynman graphs
or deal with combinatorial identities that go beyond the most
standard tree graph techniques.

The basic reason for convergence is that correlation functions of
fermions are determinants of a matrix whose entries are given by
the covariance. In contrast to the permanents of bosonic theory,
which grow factorially, such determinants are usually bounded by a
constant to the degree of the monomial that is integrated. The
"usually" is made more precise in the case where Gram bounds
apply.

It would be overly optimistic to assume that a {\em connected}
correlation function can also be represented as a determinant.
This is because a determinant is a sum over permutations, and
permutations decompose the index set into partitions (the cycles).
The sum over all permutations always contains some where the
vertex structure of the graph does not connect all different
cycles, and then the associated graph in the Feynman graph
expansion is not connected. Thus the connected correlation
functions, which receive contributions only from connected graphs,
cannot simply be determinants. However, if one takes out a minimal
connected subgraph, namely a tree, and sums over graphs that
contain this tree, one can hope to get back a determinant.
Moreover, because the number of tree graphs is much smaller than
that of all graphs, a tree sum representation is a good starting
point for a convergence proof.

One simple way to do the resummation, which we shall describe
below because it was at the beginning of this work, is a
resummation of the Mayer graphs for the generating functional of
the connected Green functions in terms of trees, in the way
proposed by Penrose a long time ago \cite{Penrose:1967}. This
provides a resummation of Feynman graphs and thus clarifies which
graphs get combined to determinants.
But although a Gram bound applies to every term in the sum over
trees, the Gram constant depends on the tree and we have no
uniform bounds for it yet (although they may be possible). The
investigation for the reasons for this problem led us in a natural
way to a positivity condition which implies uniformity, and in
fact, optimal Gram constants.

A priori, positivity plays no role in the definition of fermionic
theories (unlike the bosonic case, where only positive covariances
give well--defined Gaussian measures). In particular, fermionic
covariances of physically interesting models do not have any
positivity properties. Thus it may appear surprising that a
positivity condition plays a role in such theories. What is really
required is, however, not the positivity of the covariance but
that of a connectivity matrix that appears in the tree
representation of the connected correlation functions. It turns
out that this positivity condition is precisely equivalent to the
condition for two--body interactions that stability be preserved
in the decoupling expansion for the connected correlation
functions.

This explains, at least partially, why the applicability of Gram
bounds is fragile in that it is usually destroyed by expansion
methods that are not arranged carefully, and it also implies that
a Brydges--Battle--Federbush (BBF) representation, which preserves
stability and hence positivity, leads to uniform Gram estimates.
Indeed, it is optimal in that the Gram constant simply remains the
same as before.

We apply the Gram estimates to show norm bounds that are simple
but strong enough to study renormalization group flows and to
construct fermionic models nonperturbatively.

After finishing our proof, we discovered that Lesniewski
\cite{Lesniewski:1987} had found an explicit Gram representation
for the BBF decoupling and used it to prove cumulant bounds. The
advantage of our method is that the positivity condition makes it
obvious why the Gram bounds work by a nontechnical argument, and
thus provides a conceptually and technically simple proof, in
which details of explicit representations (such as Lesniewski's
Gram representation, which almost appears as a miracle at first
sight) are not needed.

In Sect.\ 2, we give the precise setup and state the main
analyticity theorem. Sect.\ 3 contains its proof and the formulas
for the connected correlation functions. In Sect.\ 4, we discuss
the results and some relations to other approaches.

We have written this paper so that it should be understandable for
non--experts. We assume only that the reader is familiar with
operations on finite--dimensional Grassmann algebras and some
elementary notions of graph theory, as well as the basic
connectedness formulas of polymer expansions. All these
prerequisites are well--documented, see e.g.\
\cite{Seiler:1982,Brydges:1984,Salmhofer:1999} for polymer
expansions and Appendix B of \cite{Salmhofer:1999} for Grassmann
algebras.

\section{The setup and the main result}

\subsection{The Gram Bound}
To do combinatorics with generating functions, it is convenient to
introduce a discretization, i.e.\ a space--time lattice even in
the theory with a cutoff (we shall follow the conventions of
\cite{Salmhofer:1999}). Our formulas allow us to take the
continuum limit (at fixed momentum space cutoff) in our
representation of the connected correlation functions. For the
estimates, it does not make any difference if the lattice is kept
or not.

For the moment, we just assume that the Grassmann algebra is
generated by fermionic fields $\ps(X), \psq(X)$, where $ X \in
\Xspace$ and $\Xspace$ is a finite set. A typical example for
$\Xspace$ is a lattice times a set of colour and spin indices. We
write sums over $\Xspace$ (possibly times a scaling factor, such
as a power of a lattice spacing) as $\int_\Xspace \dd X$ or
briefly $\int\dd X$ and denote by $\de_\Xspace (X,Y)$
the Kronecker delta on $\Xspace$, scaled such that 
$\int \dd X \, f(X) \de_\Xspace (X,Y)=f(Y)$. 
The Grassmann derivatives $\sfrac{\de}{\de \ps(X)}$ anticommute
and are normalized such that 
$\sfrac{\de}{\de \ps(X)} \ps(Y) = \de_\Xspace (X,Y)$.
For another family of Grassmann
variables $\et(X)$, we denote $(\ps,\et) = \int \dd X \ps(X)
\et(X)$; by the Grassmann nature of the fields, $(\ps,\et) = -
(\et,\ps)$. To a fermionic bilinear form
\begin{equation}
(\psq, C \ps) = \int \dd x \dd y \; \psq(x)\, C(x,y) \, \ps(y)
\end{equation}
we associate a Gaussian expectation value on the Grassmann algebra
by defining
\begin{equation}\label{Gauss}
\left\langle \E^{(\etq,\ps)+(\psq,\et)} \right\rangle= \int \dd
\mu_C(\ps)\; \E^{(\etq,\ps)+(\psq,\et)} = \E^{(\etq, C \et)}
\end{equation}
(the source terms being Grassmann variables as well). The elements
of the Grassmann algebra are polynomials
\begin{equation}\label{Vrep}
V(\ps) = \sli_{m,\barm \ge 0} \int \dd^{\barm} \ul{X} \dd^{m}
\ul{X}' v_{\barm,m} (\ul{X},\ul{X}') \psq^{\barm} (\ul{X}) \ps^{m}
(\ul{X}')
\end{equation}
where $\ul{X} = (X_1,\ldots X_m)$ and $\ps^m(\ul{X}) = \ps(X_1)
\ldots \ps(X_m)$. The sums over $m$ and $\barm$ are finite sums
because of the nilpotency of the Grassmann variables. The
coefficient function is chosen antisymmetric under permutation of
the $X$ variables and antisymmetric under permutation of the $X'$
variables because any other part of it would cancel out in
\Ref{Vrep}. We call $V$ even if $v_{\barm,m}=0$ unless $m+\barm$
is even
(this is in particular the case if $v_{\barm,m}=0$ unless 
$m=\barm$, but we do not need this more special condition here). 
If $V$ is even, it commutes with all other elements of
the Grassmann algebra.
Here and in the following, the notation $V(\ps)$ means that $V$ is
a polynomial in $\ps$ and $\psq$ (similarly, $\int \dd\mu_C(\ps)$
also involves integration with respect to $\psq$).

The basic reason for the convergence of fermionic perturbation
expansions is the fermionic antisymmetry. The Gaussian integral of
a monomial is zero unless there are as many $\psq$ as $\ps$ in it.
In that case, it is the determinant
\begin{equation}
\left\langle\pli_{l=1}^p \psq(Y_l)\; \pli_{k=1}^p \ps(X_k)
\right\rangle = (-1)^{p(p+1)/2} \det \big(C(X_k,Y_l)\big)_{k,l}
\end{equation}
We assume that the propagator can be written as an inner product
on some Hilbert space $\cH$, that is,
\begin{eqnarray}\label{ipro}
&& \forall X \in \Xspace \; \exists f_X,g_X\in \cH: \quad C(X,Y) =
\langle f_X\, , \; g_Y \rangle, \nonu\\ \mbox{and}&& \exists \ga_C
>0  \; \forall X \in \Xspace: \norm{f_X} \le \ga_C, \; \norm{g_X}
\le \ga_C.
\end{eqnarray}
Then the {\em Gram bound} for the determinant (see, e.g.,
\cite{Salmhofer:1999}, Appendix B.4). implies
\begin{equation}\label{Gram}
\abs{\left\langle\pli_{l=1}^p \psq(Y_l)\; \pli_{k=1}^p \ps(X_k)
\right\rangle} \le {\ga_C}^{2p}.
\end{equation}
For models of quantum field theory, a representation \Ref{ipro}
typically holds with a finite Gram constant $\ga_C$ if cutoffs are
present.

\subsection{Gaussian convolutions}
For $h>0$, we define the seminorm $\norm{V}_h$ of an element of
the Grassmann algebra given by \Ref{Vrep} by
\begin{equation}\label{normdef}
\norm{V}_h =  \sli_{m,\barm\ge 0 \atop m+\barm \ge 1} |v_{\barm, m}| h^{\barm+m}
\end{equation}
where $|v_{\barm, m}|$ is the standard norm
\begin{equation}\label{stanley}
|v_{\barm, m}| = \max_{i\in \bN_{\barm+m}}\; \sup_{X_i} \int
\pli_{j \ne i} \dd X_j \; |v_{\barm, m}(X_1,\ldots,X_{\barm+m})| .
\end{equation}
We do not assume translation invariance.
We assume that the norm of $C$ is finite: there is a constant
$\om_C$ such that
\begin{eqnarray}\label{Cdecay}
|C| &=& \max\left\{ \sup_X \int |C(X,Y)| \dd Y \;,\; \sup_X \int
|C(Y,X)|) \dd Y \right\} \nonu\\ &\le & \om_C {\ga_C}^2.
\end{eqnarray}
On the full Grassmann algebra, $\norm{\;\cdot\;}_h$ is only a
seminorm because the term $\barm=m=0$ is left out in \Ref{normdef} and
thus all constant polynomials $K$ have $\norm{K}_h=0$. On the
subspace of Grassmann polynomials with field--independ\-ent term
equal to zero, $\norm{\;\cdot\;}_h$ is a norm. The effective
action $W(V)$ defined below is defined such that $W(V)(0)=0$, so
it is in that subspace.

Let
\begin{equation}
U(\ph) = (\mu_C* V)(\ph) = \int\dd\mu_C (\ps) V(\ps+\ph)
\end{equation}
be the convolution of $V$ with $\mu_C$. $U$ has an expansion of
type \Ref{Vrep}, with
\begin{eqnarray}\label{practicallynosign}
u_{\barm, m}(\ul{Y},\ul{Y}') &=& \sli_{\bar n \ge \barm}
\tst{{\bar n \choose \barm}} \sli_{n \ge m} \tst{{n \choose m}}
\int \dd^{\bar n-\barm} \ul{X} \int \dd^{n-m} \ul{X}
\\
&& (-1)^{m(\bar n - \barm)} v_{\bar n, n}
(\ul{Y},\ul{X},\ul{Y}',\ul{X}') \left\langle \psq^{\bar n-\barm}
(\ul{X}) \ps^{n-m} (\ul{X}') \right\rangle. \nonu
\end{eqnarray}
Here we used the antisymmetry of the coefficient $v_{\barm, m}$.
Taking the norm gives, by the Gram estimate \Ref{Gram},
\begin{equation}
|u_{\barm, m}(\ul{Y},\ul{Y}')| \le \sli_{n \ge m} {n \choose m}
\sli_{\bar n \ge \barm} {\bar n \choose \barm} |v_{\bar n,n}|
{\ga_C}^{n-m+\bar n -\barm},
\end{equation}
so
\begin{equation}\label{convonorm}
\norm{\mu_C* V}_h \le \norm{V}_{h+\ga_C}.
\end{equation}
Thus integrating over fermionic variables only shifts the norm
parameter by the Gram constant. This is in strong contrast to
bosonic problems, and a first indication for the convergence of
perturbation theory for fermions. The above estimate is, however,
not sufficient because it does not lead to bounds that are uniform
in $|\Xspace|$. For this we need to assume decay of the covariance
and of the $v_m$, and consider connected functions, such as, e.g.,
generated by the effective action.

\subsection{A norm bound for the effective action}
Let $V$ be even. We define the effective action as
\begin{equation}
W(V) (\ps) = \log \frac{1}{Z}\int \dd \mu_C(\ps') \;
\E^{V(\ps+\ps')} = \left(\log\frac{1}{Z} \mu_C * \E^{V}\right)
(\ps)
\end{equation}
where $Z=\mu_C * \E^{V}\vert_{\ps=0}$, so that $W(V)(0)=0$. For
finite $\Xspace$, the argument of the logarithm,
$\sfrac{1}{Z}\mu_C * \E^{V} $, is a polynomial whose constant term
is $1$; thus $W(\ps)$ is well-defined if $\norm{V}_h$ is small
enough (depending on $\Xspace$) because the expansion for the
logarithm terminates after a finite number of terms by nilpotency
of the Grassmann variables. The following theorem implies that
analyticity holds uniformly in $|\Xspace|$ provided the Gram
constant $\ga_C$ and the decay constant $\om_C$ are bounded
uniformly in $|\Xspace|$.

\begin{theorem} \label{th1}
Assume \Ref{ipro} and \Ref{Cdecay}. Let $V$ be even, $V(0)=0$, and
$h'=h+3\ga_C$. If $ {\om_C} \; \norm{V}_{h'} < 1$ then $W$ is
analytic in $V$ and in the fields, and
\begin{equation}\label{logbo}
\norm{W}_h \le - \frac{1}{{\om_C}} \log \left( 1-{\om_C}\;
\norm{V}_{h'}\right).
\end{equation}
Let $W(V) = \sum_{p \ge 1} W_p(V)/p!$ be the expansion of $W$ in
powers of $V$. Then for all $P \ge 1$,
\begin{equation}\label{orderP}
\norm{W(V) - \sli_{p=1}^{P} \frac{1}{p!} W_p(V)}_h
\le
{\om_C}^P \frac{{\norm{V}_{h'}}^{P+1}}{1- \om_C \norm{V}_{h'}}.
\end{equation}
Moreover, we can replace $h'$ by $h''=h+2\ga_C$ in \Ref{logbo} and
\Ref{orderP} if we replace $\om_C$ by $2 \om_C$ in these bounds.
\end{theorem}
In particular, for $P=1$,
\begin{equation}
\norm{W(V)- \mu_C*V}_h \le
{\om_C} \; \frac{{\norm{V}_{h'}}^2}%
{1-{\om_C}\,\norm{V}_{h'}}.
\end{equation}
A difference to the linear estimate \Ref{convonorm} is that the
shift in the norm parameter $h$ is not $\ga_C$ but $\be \ga_C$
with $\be > 1$. It will be explicit in the proof where this factor
comes from; the last statement of the theorem shows that there is
some freedom in moving factors around in the constants. However,
we have not been able to prove a bound where the norm parameter
shifts only by $\ga_C$.

We shall discuss a related bound in Section \ref{cumusect}.

\section{The expansion for the effective action}

\subsection{Connected parts and logarithms}

In this section, we briefly recall a characterization of connected
parts and their role in taking logarithms.
Let $\bN_p = \nato{p}$, let $\cA $ be a commutative algebra with
unit $1$, and assume a function
\begin{equation}
\al: \bP (\bN_p) \to \cA, \quad Q \mapsto \al(Q)
\end{equation}
with $\al(\emptyset) =1$ to be given (here $\bP (M)$ is the power
set of $M$).

\begin{lemma} \label{uniconn}
There is a unique function
\begin{equation}
\connpart{\al}: \bP (\bN_p) \to \cA, \quad Q \mapsto
\connpart{\al}(Q), \quad \connpart{\al}(\emptyset)=0,
\end{equation}
that satisfies
\begin{equation}\label{coneq}
\forall Q \subset \bN_p : \al(Q) = \sli_{J_0 \subset Q \atop \min
Q \in J_0} \connpart{\al}(J_0) \; \al(Q \setminus J_0).
\end{equation}
Moreover, $\al(Q)$ is the sum over partitions of $Q$ of products
of $\connpart{\al}$ of the elements of the partition (here
$\dotcup$ denotes the disjoint union):
\begin{equation} \label{partition}
\al(Q) = \sli_{m \ge 1} \frac{1}{m!} \sli_{I_1,\ldots I_m \ne
\emptyset \atop I_1 \dotcup \ldots \dotcup I_m = Q} \;
\pli_{l=1}^m \connpart{\al} (I_l)
\end{equation}
\end{lemma}

\begin{proof}
Induction on $|Q|$ gives existence and uniqueness of
$\connpart{\al}$: For $|Q|=1$, \Ref{coneq} is simply
$\connpart{\al}(Q)=\al(Q)$. Once $\connpart{\al}(Q')$ has been
determined for all $Q'$ with $|Q'| < |Q|$, \Ref{coneq} is solved
in the form
\begin{equation}
\connpart{\al}(Q) = \al(Q) - \sli_{J_0 \subset Q \atop \min Q \in
J_0 \ne Q} \connpart{\al}(J_0) \; \al(Q \setminus J_0)
\end{equation}
The right hand side of \Ref{partition} solves \Ref{coneq}.
\end{proof}

\noin 
The convention $\connpart{\al}(\emptyset)=0$ has no consequences
because $\connpart{\al}(\emptyset)$ never appears in any formula.

\begin{lemma} \label{takethelog}
As a formal series in $\connpart{\al}$,
\begin{equation} \label{polylog}
\log \big(1+ \sli_{Q \subset \bN_p \atop Q \ne \emptyset} \al(Q)
\big) = \sli_{m \ge 1} \frac{1}{m!} \sli_{I_1,\ldots I_m \subset
\bN_p \atop \mathrm{all \; nonempty}} \connpart{\cU}^{(m)}
(I_1,...,I_m) \pli_{l=1}^m \connpart{\al} (I_l)
\end{equation}
where $\connpart{\cU}^{(1)}=1$ and for $m \ge 2$,
\begin{equation}
\connpart{\cU}^{(m)} (I_1,\ldots,I_m) = \sli_{G \in
\connpart{\cG}(\bN_m)} \pli_{(i,j) \in G} \ga(I_i,I_j)
\end{equation}
with $\ga(I_i,I_j) = -1 $ if $I_i \cap I_j \ne \emptyset$ and $0$
otherwise, and $\connpart{\cG}(\bN_p)$ the set of connected graphs
on $\bN_p$.
In particular, if $\la_1,\ldots, \la_p$ are formal parameters,
then
\begin{equation}\label{laset0}
\frac{\del^p}{\del \la_1 \ldots \del \la_p} \log \big(1+ \sli_{Q
\subset \bN_p \atop Q \ne \emptyset} \al(Q) \pli_{q \in Q} \la_q
\big) \Big\vert_{\la=0} = \connpart{\al} (\bN_p) .
\end{equation}
\end{lemma}

\begin{proof}
When \Ref{partition} is inserted to replace $\al(Q)$, $\ze = 1+
\sum \al(Q)$ takes the form of a polymer partition function, with
the nonempty subsets of $\bN_p$ as polymers and disjointness as
the compatibility relation. Eq.\ \Ref{polylog} is the standard
polymer formula for the logarithm of the partition function
\cite{Seiler:1982,Salmhofer:1999}. Eq.\ \Ref{laset0} follows by
noting that for all $m \ge 2$, the connectedness condition in the
function $\connpart{\cU}^{(m)}$ implies that after
differentiation, some factors $\la_i$ remain, so that evaluating
at zero picks out the term $m=1$ from the sum (see
\cite{Salmhofer:1999}, Section 2.5).
\end{proof}

\subsection{Connected parts of the Laplacian}
We expand the effective action
\begin{equation}\label{Wpexp}
W(\la V) = %\log \left( \E^{\De} \E^{\la V} \right) =
\sli_{p \ge 1} \frac{\la^p}{p!} W_p(V)
\end{equation}
with $W_p(V) = \truncated{\langle V;\ldots;V \rangle} -
\sfrac{\del^p}{\del\la^p} \log Z\mid_{\la=0}$, where, for elements
$V_1, \ldots, V_p$ of the even subalgebra,
\begin{equation}
\truncated{\langle V;\ldots;V \rangle} = \left[
\sfrac{\del^p}{\del\la_1 \ldots \del\la_p}
%\left(
\log \left( %%%\frac{1}{Z }
\mu_C * \E^{\la_1 V_1 + \ldots + \la_p V_p}\right)
%-\log Z \right)
\right]_{\la_q=0 \;\forall q}  %%%_{\la_1=\ldots=\la_p=0}
\end{equation}
is the connected correlation function of $V_1, \ldots, V_p$. It is
an element of the even subalgebra. The subtraction of $\log Z$
removes the $\psi$--independent term from $W_p(V)$. Because the
derivative is evaluated at $\la =0$, we can replace $\mu_C *
\E^{\la_1 V_1 + \ldots + \la_p V_p}$ by
\begin{equation}\label{krimskrams}
\mu_C * \pli_{q=1}^p (1+ \la_q V_q) = 1+ \sli_{Q \subset \bN_p
\atop Q \ne \emptyset} \al(Q) \; \pli_{q\in Q} \la_q
\end{equation}
with
\begin{equation}
\al(Q) = \mu_C * \pli_{q\in Q} V_q.
\end{equation}
Similarly, we can replace $Z$ by \Ref{krimskrams} evaluated at
$\ps=0$. Because all $V_q$ are in the even subalgebra, $\al(Q)$ is
in the even subalgebra, and hence all $\al$'s commute. Thus, by
Lemma \ref{takethelog},
\begin{equation}
\truncated{\langle V;\ldots;V \rangle} = \connpart{\al} (\bN_p).
\end{equation}
We now rewrite Gaussian convolutions in terms of the action of a
Laplacian acting on $p$ independent copies of the field $\ps$;
this is convenient for doing the combinatorics.

\begin{lemma}\label{jomei}
Let $\De = \sli_{q,q'=1}^p \De_{qq'}$ where
\begin{equation}\label{Deqqdef}
\De_{qq'} = - \left( \frac{\de}{\de \ps_q}, C \frac{\de}{\de
\psq_{q'}}\right)
\end{equation}
Then
\begin{equation}
\label{eding} \left( \mu_C * \pli_{q\in Q} V_q \right) (\psq,\ps)
=
\left[ \E^{\De} \pli_{q\in Q} V_q (\psq_q,\ps_q)
\right]_{{\psq_q=\psq \atop \ps_q=\ps}\; \forall q \in \bN_p}
\end{equation}
\end{lemma}

\begin{proof}
See Appendix \ref{detstuff}.
\end{proof}

\noin Because the exponential of the Laplacian acts on a product
over $q \in Q$, it suffices to get an expression for the connected
part of $\E^{\De}$:
\begin{equation}
\truncated{\langle V_1;\ldots;V_p \rangle} = \connpart{(\E^{\De})}
\pli_{q=1}^p V_q
\end{equation}

\subsection{A direct resummation}
We now discuss a representation of the connected part of the
Laplacian as a sum over trees which corresponds to a direct
resummation of the Feynman graph expansion, to motivate the
solution to the problem. 
%Readers that are already motivated can skip this section.

Because all $\De_{qq'}$ commute with one another,
\begin{eqnarray}
\E^{\De} &=& \pli_{q=1}^p \E^{\De_{qq}} \pli_{q < q'} (1+
\E^{\De_{qq'}+\De_{q'q}} -1) \nonu\\ &=& \pli_{q=1}^p
\E^{\De_{qq}} \sli_{G \in \cG(\bN_p)} \pli_{\{q,q'\} \in G}
(\E^{\De_{qq'}+\De_{q'q}} -1)
\end{eqnarray}
with $G$ summed over all graphs on $\bN_p$ (that is, the set of
all subsets of $\bN_p$ that have two elements). Decomposing every
$G$ into its connected components, we get (cf.\ \Ref{partition})
\begin{equation} \label{Gcgen}
\connpart{(\E^{\De})} = \pli_{q=1}^p \E^{\De_{qq}} \sli_{G \in
\connpart{\cG}(\bN_p)} \pli_{\{q,q'\} \in G}
(\E^{\De_{qq'}+\De_{q'q}} -1)
\end{equation}
with $G$ now summed over connected graphs on $\bN_p$
\cite{Seiler:1982,Salmhofer:1999}.

Applying $\connpart{(\E^\De)}$ to $\prod V_q$ generates the
expansion of $\truncated{\langle V;\ldots;V \rangle}$ as a sum of
values of Feynman graphs. Because every $G$ is connected, all
Feynman graphs that contribute are connected (see Sections 2.3 and
2.4 of \cite{Salmhofer:1999}). 

A term by term estimation of the sum in \Ref{Gcgen} cannot lead to
convergence: $|\cG(\bN_p)|=2^{p\choose 2}$ % \sim \E^{{\log2 \over 2} p^2}$ 
and $\cG_c$ is of similar size. 
The $1/p!$ in \Ref{Wpexp} %is only $\sim \E^{-p\log p}$, 
decreases more slowly, 
so the majorant series obtained by term--by--term estimation diverges.

However, one can partly resum the expansion, to get a sum over
trees (connected graphs without loops), using

\begin{lemma}\label{lpen}
To every tree $T \in \cT(\bN_p)$ there is a graph $H^*(T) \in
\cG(\bN_p)$ such that $T \cap H^*(T)= \emptyset$ and
$\cG_c(\bN_p)$ is the disjoint union
\begin{equation}
\cG_c(\bN_p) = \bigcup_{T \in \cT} \{H \cup T: H \subset H^*(T)\}.
\end{equation}
\end{lemma}

\noin We include Penrose's proof \cite{Penrose:1967} in Appendix
\ref{penapp}. Another proof can be found in \cite{Seiler:1982}. An
immediate consequence of Lemma \ref{lpen} is a representation of
the connected  correlations as a sum over trees.

\begin{theorem}
Let $\De^{(T)} = \sli_{q=1}^p \De_{qq} + \sli_{\{q,q'\} \in
H^*(T)} (\De_{qq'}+\De_{q'q}) $. Then
\begin{equation} \label{nicetry}
\connpart{(\E^{\De})} = \sli_{T \in \cT (\bN_p)} \E^{\De^{(T)}}\;
\pli_{\{q,q'\} \in T} (\E^{\De_{qq'}+\De_{q'q}} -1).
\end{equation}
\end{theorem}

\begin{proof}
Let $a_{qq'}=\E^{\De_{qq'}+\De_{q'q}}-1$.
By Lemma \ref{lpen} and the binomial theorem,
\begin{eqnarray}
%&& 
\sli_{G \in \cG_c(\bN_p)} \pli_{\{q,q'\} \in G} a_{qq'} %\nonu\\
&=& \sli_{T\in \cT(\bN_p)} \pli_{\{q,q'\} \in T}  a_{qq'} \sli_{H
\subset H^*(T)} \pli_{\{q,q'\} \in  H}  a_{qq'}
\\
&=& \sli_{T\in \cT(\bN_p)} \pli_{\{q,q'\} \in  T} a_{qq'}
\pli_{\{q,q'\} \in  H^*(T)} \E^{\De_{qq'}+\De_{q'q}} \nonu
\end{eqnarray}
\end{proof}

\noindent Finally, we can use $\E^\De-1 = \De \int_0^1 \dd s\,
\E^{s\De}$ on every line of the tree, to get
\begin{equation}\label{Schmarrn}
\connpart{(\E^{\De})} = \sli_{T \in \cT (\bN_p)} \pli_{\{q,q'\}
\in T} ({\De_{qq'}+\De_{q'q}})\; \int \dd\sss\,
\E^{\De^{(T,\sss)}}\;
\end{equation}
with $\sss = (s_\ell)_{\ell \in T}$, $\dd\sss = \prod_{\ell \in T}
\dd s_\ell$, and
\begin{equation} \label{Kompott}
\De^{(T,\sss)} = \De^{(T)} + \sli_{\{q,q'\} \in T}
s_{\{q,q'\}}({\De_{qq'}+\De_{q'q}}).
\end{equation}
By Cayley's theorem,
\begin{equation}
|\cT(\bN_p)| \le p^{p-2} \le (p-1)!\; \E^{p-1},
\end{equation}
and the $(p-1)!$ gets cancelled by the $p!$ in the denominator in
\Ref{Wpexp}. Thus this resummation will lead to a convergence
proof if the action of each summand on $\prod V_q$ can be bounded
uniformly in $T$. It is at this point that a problem arises with
the representation \Ref{Schmarrn}. The action of
$\E^{\De^{(T,\sss)}}$ on a monomial gives a determinant, but we
have no bound for the corresponding Gram constant that is uniform
in $T$. We explain the reasons for this in the following and then
derive a representation that looks very similar to \Ref{Schmarrn},
but leads to uniform Gram constants.

\subsection{Positivity and Gram estimates}

A general feature of tree expansions like \Ref{Schmarrn} is that
the Laplacians appearing in $\connpart{(\E^\De)}$ depend on the
tree $T$ and further parameters. To discuss this dependence, we
introduce the following notation. For a matrix $M \in M_p (\bC)$
and $Q \subset \bN_p$, let
\begin{equation}
\De_Q[M] = \sli_{q,q'\in Q} M_{qq'} \De_{qq'}.
\end{equation}
We abbreviate $\De_{\bN_p} [M] = \De[M]$. 
The matrix element
$M_{qq'}$ can be thought of as a weight factor associated to the 
directed line $(q,q')$. The matrices $M$ occurring in our
Laplacians will always be real and symmetric.
The Laplacian acting in
\Ref{eding} is $\De[P]$, where $P_{qq'}=1$ for all $q$ and $q'$.
Note that
$P$ is $p$ times the orthogonal projection to the space spanned by
the vector $(1,\ldots,1)$, so $P$ is positive 
({\em we call a matrix $M$ positive, 
and write $M \ge 0$, if 
$M$ is hermitian and has nonnegative eigenvalues}). 
It is the positivity of the coefficient matrix $M$ which will
be crucial for good estimates. The structure of the matrices $M$
belonging to the Laplacians in \Ref{Schmarrn} is discussed in
detail in Appendix \ref{netpos}.

The product $V_1(\ps_1) \ldots V_p (\ps_p)$ is linear in every
factor, so we can for the following restrict to a single summand
$v_{\barm_q, m_q}$ from the representation \Ref{Vrep} for every
$q$. Thus the Laplacian now acts on an element of degree $\barm_1
+ \ldots + \barm_p$ in $\psq$ and $m_1 + \ldots + m_p$ in the
$\ps$. It will be convenient to keep the coefficient function
$v_{\barm_q, m_q}$ and the integral over the $X$ variables.

Let $\Bspace=\bN_p \times \Xspace$, and for $\bb = (q,X) \in
\Bspace$ let $\Ps(\bb) = \ps_q(X)$ and $\Psq(\bb) = \psq_q (X)$.
Introducing
\begin{equation}\label{CCdef}
\CC ((q,X),(q',X')) = M_{qq'} C(X,X'),
\end{equation}
and using the notation $\int_\Bspace \dd \xi F(\xi) = \sum_{q=1}^p
\int_\Xspace \dd X F(q,X)$, we have
\begin{equation}
\De[M] = - \ili_\Bspace \dd \bb \ili_\Bspace \dd \bb' \;
\frac{\de}{\de\Ps(\bb)}  \CC (\bb,\bb') \frac{\de}{\de\Psq(\bb')}
=\DDe_\CC .
\end{equation}
Then
\begin{equation}
\E^{\De[M]} \pli_{q=1}^p {\ps_q}^{m_q} (\ul{X}_q) \; \pli_{q=1}^p
{\psq_q}^{\barm_q} (\ul{X}'_q)
=
\E^{\DDe_\CC} \pli_{\bb \in D} \Ps(\bb) \pli_{\bb' \in \bar D}
\Psq(\bb')
\end{equation}
where $D \subset \Bspace$ and $\bar D \subset \Bspace$ are
determined by the $\ul{X}_q$ and $\ul{X'}_q$.

For subsets $A, \bar A$ of $\Bspace$ with $|\bar A| = |A| = d$,
denote the corresponding minor of $\CC$ by $\CC_{\bar A,A}$, that
is, if we order $\Bspace $ in some way, and if $A=\{a_1, \ldots,
a_d\}$ with $a_1 < \ldots < a_d$ and $\bar A=\{\bar a_1, \ldots,
\bar a_d\}$ with $\bar a_1 < \ldots < \bar a_d$, then $\CC_{\bar
A,A}$ is the $d \times d$ matrix with entries
\begin{equation}
\left(\CC_{\bar A,A}\right)_{i,j} = \CC_{\bar a_i,a_j}.
\end{equation}

\begin{lemma} \label{lapdet}
There are $\veps_{D\bar D}^{A \bar A} \in \{1,-1\}$ such that
\begin{eqnarray}
&& \E^{\DDe_\CC} \pli_{\bb \in D} \Ps(\bb) \pli_{\bb' \in \bar D}
\Psq(\bb')= \nonu\\ && \sli_{A \subset D, \bar A \subset \bar D
\atop |A| = |\bar A|} \veps_{D\bar D}^{A \bar A} \; \det
\left(\CC_{A, \bar A}\right) \pli_{\bb\in D\setminus A} \Ps_\bb
\pli_{\bb'\in \bar D \setminus \bar A} \Psq_\bb.
\end{eqnarray}
\end{lemma}

\begin{proof}
Expand and permute. 
\end{proof}

\noin Thus we have to estimate determinants. The point is now that
good Gram estimates require some positivity.

We call a matrix $A$ a {\em Gram matrix with Gram constant $\al$}
if there is a Hilbert space $\cH$ and there are vectors $f_i$ and
$g_j$  with $\norm{f_i} \le \al$ and $\norm{g_j } \le \al$ such
that $A_{ij} = \langle f_i, g_j \rangle$.

\begin{lemma} \label{lelem}
If $A$ is a Gram matrix with Gram constant $\al$, then every minor
$A_{\bar D,D}$ is a Gram matrix with Gram constant $\al$, and
\begin{equation} \label{gramin}
\abs{\det A_{\bar D, D}} \le \al^{|D|+|\bar D|} = \al^{2|D|}.
\end{equation}
If $A$ and $B$ are Gram matrices with Gram constants $\al$ and
$\be$, and if $C_{ij} = A_{ij} B_{ij}$, then $C$ is a Gram matrix
with Gram constant $\al\be$.
\end{lemma}
\begin{proof}
The statement about minors is trivial; \Ref{gramin} is Gram's
inequality (see e.g.\ Appendix B.4 of \cite{Salmhofer:1999}). If
$A_{ij}=\langle a_i, \tilde a_j \rangle$ and $B_{ij} = \langle
b_i, \tilde b_j \rangle$, then $C_{ij} = \langle a_i\otimes b_i,
\tilde a_j\otimes \tilde b_j \rangle$ is also a Gram matrix, with
Gram constant $\al\be$.
\end{proof}

\noin Every nonnegative matrix is a Gram matrix:
\begin{lemma} \label{goodguy}
Let $A$ be a real matrix, $A=A^T$, $A \ge 0$ (that is, all
eigenvalues of $A$ are nonnegative). Then $A$ is a Gram matrix and
\begin{equation} \label{BB}
0 \le \det A \le \pli_{i=1}^n A_{ii}.
\end{equation}
\end{lemma}
\begin{proof}
All eigenvalues of $A$ are nonnegative, so there is a real matrix
$B$, with $B=B^T \ge 0$ such that $A=BB^T =B^2$. If $b_i=
(b_{ik})_k$ is the $i^{\rm th}$ row vector of $B$, this means
$A_{ij} = \langle b_i, b_j \rangle$, thus in particular $A_{ii} =
\norm{b_i}^2$. The Gram inequality implies $\det A \le \prod_i \norm{b_i}^2$, 
so \Ref{BB} holds.
\end{proof}

\noin If $A=A^T$, but $A$ is not necessarily positive, $A$ can
also be written as a Gram matrix by the polar decomposition.
However, now $A_{ij} = \langle \tilde b_i, b_j \rangle$, and
instead of an equality, one only has $A_{ii} = \langle \tilde b_i,
b_i \rangle \le \Vert\tilde b_i\Vert \norm{b_i}$, so the Gram
bound for the determinant is not just a bound by the product of
the diagonal elements. In general, it is not easy to get bounds on
the norm of the $b_i$ and $\tilde b_i$. The absence of these
bounds is exactly the problem with the tree representation
\Ref{Schmarrn} of the connected correlations (we discuss this in
Appendix \ref{netpos}).

But Lemma \ref{goodguy} also suggests a way out of this problem.
The classic Brydges--Battle--Federbush interpolation that
preserves stability of potentials will, as we shall see, also
preserve the positivity of the matrix $M$ that appears in
$\De[M]$. The following immediate consequence of Lemma
\ref{goodguy} then implies uniformity of the Gram constant.

\begin{lemma}\label{thatsit}
Let $M$ be real and symmetric, and $M\ge 0$, with diagonal
elements $M_{qq} \le 1$ for all $q \in \bN_p$. Assume \Ref{ipro}.
Then $\CC$, given by \Ref{CCdef}, is a Gram matrix with Gram
constant $\ga_C$.
\end{lemma}

\begin{proof}
By Lemma \ref{goodguy}, $M$ is a Gram matrix with Gram constant
$1$. Let $M_{qq'} = \langle b_q, b_{q'} \rangle$ be its Gram
representation. By \Ref{ipro},
\begin{equation}
\Ga ((q,X),(q',X')) = \langle b_q \otimes f_X, b_{q'} \otimes
g_{X'} \rangle.
\end{equation}
As in the proof of Lemma \ref{lelem}, the Gram bound implies the
statement.
\end{proof}

\noin The matrix $P$ appearing in the Laplacian in \Ref{eding} is
a positive multiple of a projection, so $P \ge 0$. Moreover, all
diagonal elements of $P$ are equal to $1$. Decoupling
off--diagonal blocks preserves these properties:

\begin{lemma} \label{interposo}
For $M \in M_p(\bR)$, $s \in [0,1]$, and $A \subset \bN_p$, let
$\left(M^{(A,s)}\right)_{q,q'} = s M_{qq'}$ if $q\not\in A$ and
$q'\in A$, or if $q\in A$ and $q'\not\in A$, and
$\left(M^{(A,s)}\right)_{q,q'} = M_{qq'}$ otherwise.
%
%    keep this in case the macro conflicts can be fixed!
%
%\begin{equation}
%\left(M^{(A,s)}\right)_{q,q'} =
%\cases{ s M_{qq'} & if $q\in A$ and $q'\not\in A$ \cr
%& or $q\not\in A$ and $q'\in A$ \cr
%M_{qq'} & otherwise.}
%\end{equation}
%
Then the diagonal elements of $M^{(A,s)}$ remain unchanged,
\begin{equation}
\left(M^{(A,s)}\right)_{qq} = M_{qq} \quad \forall q \in \bN_p,
\end{equation}
and if $M=M^T \ge 0$, then the same holds for $M^{(A,s)}$.
\end{lemma}

\begin{proof}
It is obvious that the diagonal elements remain unchanged and that
the matrix remains symmetric.
By permuting the rows and columns of $M$ with the same
permutation, which amounts to a change of basis and therefore does
not change positivity properties, we can assume that $A=\bN_r$ for
some $r \le p$, and thus get, with $A^c=\bN_p \setminus A$,
\begin{equation}
M^{(A,s)} = \left(\begin{array}{cc} M_{AA} & s M_{AA^c} \\
sM_{AA^c}^T & M_{A^cA^c}
\end{array} \right)
=
s M + (1-s) \left(\begin{array}{cc} M_{AA} & 0 \\ 0 & M_{A^cA^c}
\end{array} \right).
\end{equation}
The blockdiagonal matrix inherits positivity from $M$. Thus
$M^{(A,s)}$ is a convex combination of two positive matrices,
hence positive.
\end{proof}

\noin A tree expansion leading to uniform Gram constants is given
in the following theorem.

\begin{theorem}\label{BBFth}
Let $M$ be a real symmetric matrix, and $M \ge 0$.
Then
\begin{eqnarray}\label{BBFrep}
\connpart{\left( \E^{\De[M]}\right)} (\bN_p) &=& \sli_{T \in
\cT(\bN_p)} \pli_{\{q,q'\} \in T} M_{qq'} (\De_{qq'} + \De_{q'q})
\nonu\\ && \ili_{[0,1]^{p-1}} \dd\sss \;
\sli_{\pi \in \Pi(T)} \vphi(T,\pi,\sss) \E^{\De[M(T,\pi,\sss)]}
\end{eqnarray}
where $\sss = (s_1, \ldots, s_{p-1})$, $\dd\sss =\dd s_1 \ldots
\dd s_{p-1}$, $\vphi(\pi,\sss) \ge 0$, and $M(T,\pi,\sss)$ is a
nonnegative symmetric matrix with diagonal entries
$\big(M(T,\pi,\sss)\big)_{qq}=M_{qq}$. The sum over $\pi$ runs
over a $T$--dependent set $\Pi(T)$ of permutations $\pi$ of
$\bN_p$, and
\begin{equation} \label{normalized}
\int \dd\sss  \;
\sli_{\pi \in \Pi(T)} \vphi(T,\pi,\sss) =1.
\end{equation}
\end{theorem}

\medskip\noin
This is a variant of the BBF formula
\cite{Battle/Federbush,Brydges:1984,Glimm/Jaffe:1987,Mack/Pordt}.
It is proven by a repeated application of Lemma \ref{interposo}.
We include a simple proof of Theorem \ref{BBFth}, which avoids all
explicit details about $\vphi(T,\pi,\sss)$ that we are not going
to need, in Appendix \ref{BBFapp}. The essential points we need,
namely the positivity of $M(T,\pi,\sss)$ and \Ref{normalized}, do
not depend on these details.

\subsection{Proof of Theorem \ref{th1}}
By Theorem \ref{BBFth} and Lemma \ref{thatsit}, we can now bound
$\norm{\truncated{\langle V;\ldots;V \rangle}}_h$ essentially by a
sum over trees, to which standard procedures apply, as follows.
The action of $\prod_{\{q,q'\} \in T} (\De_{qq'}+\De_{q'q})$
on the homogeneous polynomial 
\begin{equation}
\int \pli_{q=1}^p \dd \ul{X}_q \dd \ul{X}_q' \; v_{\barm_q,m_q}
(\ul{X}_q,\ul{X}_q') \psq_q^{\barm_q}(\ul{X}_q) \ps_q^{m_q}
(\ul{X}_q')
\end{equation}
is as follows. Let the tree $T$ have incidence numbers $d_1,
\ldots d_p$. Then  $d_q=\th_q+\bar \th_q$ derivatives act on the
$q^{\rm th}$ factor, $\th_q$ of them with respect to $\ps_q$, and
$\bar\th_q$ with respect to $\psq_q$. Because the coefficient
function is totally antisymmetric, these derivatives gives rise to
a combinatorial factor $\barm_q (\barm_q-1) \ldots (\barm_q-\bar
\th_q+1)$ $m_q (m_q-1) \ldots (m_q-\th_q+1)$, that is,
\begin{equation}
{\barm_q \choose \bar \th_q} \bar\th_q ! \; {m_q \choose \th_q}
\th_q! ,
\end{equation}
times a monomial of total degree $m_q-\th_q+\barm_q-\bar\th_q$ for
every $q$. Applying $\E^{\De[M(T,\pi,\sss)]}$ to the product of
these monomials gives, by Lemma \ref{lapdet}, a sum over subsets
$A, \bar A$ of determinants of minors determined by $A$ and $\bar
A$ (these subsets are unions of subsets $\bar A_q$ and $A_q$ for
every factor belonging to $q \in \bN_p$). Estimate the
determinants. Because $M(T,\pi,\sss)$ is positive and has diagonal
elements bounded by $1$, the Gram constant of the corresponding
matrices $\CC$ is $\ga_C$ independent of $T,\sss$, and $\pi$.
Thus, by Lemma \ref{thatsit}, each determinant is bounded by
${\ga_C}^{a_q+\bar a_q}$, where $a_q=|A_q|$, $\bar a_q=|\bar
A_q|$. We use \Ref{normalized} to do the $\sss$--integral and the
sum over $\pi$. By Cayley's theorem on the number of trees with
fixed incidence numbers $d_1, \ldots, d_p$ (see, e.g.\
\cite{Glimm/Jaffe:1987}, Section 20.3), we can sum over incidence
numbers, and are left with
\begin{eqnarray}\label{lengthy}
\norm{W_p(V)}_h &\le& \sli_{m_1, \ldots, m_p \ge 1} \sli_{\barm_1,
\ldots, \barm_p \ge 1} \cS \left( (m_q,\barm_q)_{q \in
\bN_p}\right)
\\
&\sli_{d_1,\ldots d_p \ge 1 \atop d_1+... +d_p = 2(p-1)} &
\sfrac{(p-2)!}{(d_1-1)! \ldots (d_p-1)!}
\sli_{\th_1,\ldots,\th_p,  \bar \th_1, \ldots, \bar \th_p \ge 0
\atop \th_q+\bar\th_q=d_q, \th_q \le m_q, \bar\th_q \le \barm_q}
\tst{{\barm_q \choose \bar \th_q} \bar\th_q ! \; {m_q \choose
\th_q} \th_q!} \nonu\\
&\sli_{a_1,\ldots,a_p \ge 0 \atop \bar a_1, \ldots, \bar a_p \ge
0} & \pli_{q=1}^p \tst{{m_q -\th_q \choose a_q} {\barm_q-\bar
\th_q \choose \bar a_q}} \;
h^{m_q-\th_q-a_q+\barm_q-\bar\th_q-\bar a_q} {\ga_C}^{a_q+\bar
a_q} . \nonu
\end{eqnarray}
The binomials come from the number of subsets $A_q$ with
$|A_q|=a_q$, and
\begin{eqnarray}\label{cSdef}
\cS &=& \sup\limits_{T \in \cT(\bN_p)} 
\sup\limits_{\tilde X}
\max\limits_{i}
\int \dd\ul{X} \dd\ul{X}'
\int \dd\ul{Y} \dd\ul{Y}' \;
\de_\Xspace (\tilde X,Z_i)
\nonu\\
&&\pli_{\{q,q'\} \in T} \AbsC(X_q,X'_{q'})\;
\pli_{q=1}^p |v^{(q)}_{\barm_q, m_q}
(\ul{X},\ul{Y},\ul{X}',\ul{Y}')|
\end{eqnarray}
with
\begin{equation}
\AbsC(X,X') = \max \{ |C(X,X')|, |C(X',X)|\},
\end{equation}
and where $Z_i$ denotes one of the coordinates in $\ul{X},\ul{Y},\ul{X}',\ul{Y}'$ that is fixed to $\tilde X$ by the
delta function. The supremum over $\tilde X$ is the supremum in the 
definition \Ref{stanley} of $\norm{\;\cdot\;}_h$. 
(here we used that in the seminorm $\abs{\;\cdot\;}_h$, 
the field--independent term is left out. 
We give a bound for this term %(the cumulant) 
in Section \ref{cumusect}). Root the tree at the $q$ for
which $\tilde X$ appears as an argument of $v^{(q)}$, 
and perform the integrals in \Ref{cSdef} by trimming the tree in
the usual way (see \cite{Brydges:1984}, Appendix C) and using the
summability \Ref{Cdecay} of the propagator. This gives
\begin{equation}
\cS  \le |C|^{p-1} \pli_{q=1}^p |v^{(q)}_{\barm_q, m_q}|
\le {\om_C}^{p-1}\; {\ga_C}^{2(p-1)} \pli_{q=1}^p
|v^{(q)}_{\barm_q, m_q}| .
\end{equation}
The sums over $a_q$ and $\bar a_q$ give
$(h+\ga_C)^{\barm_q-\bar\th_q+m_q-\th_q}$. The incidence numbers
$d_q$ on the tree satisfy
\begin{equation}\label{inzi}
2(p-1) = \sli_{q=1}^p d_q = \sli_{q=1}^p (\th_q + \bar \th_q).
\end{equation}
Because $\bar\th_q+\th_q=d_q \ge 1$,
\begin{equation}
\frac{\th_q! \bar \th_q!}{(d_q-1)!} \le \max\{\th_q,\bar\th_q\}.
\end{equation}
Using this bound we can sum over the $d_q$ (dropping the
constraint that $d_1+ \ldots +d_p = 2(p-1)$) and thus remove the
constraint $\bar\th_q+\th_q=d_q$ in the $\th$ sums. The remaining
sums over the $\th_q$ and $\bar \th_q$ are bounded by
\begin{equation}\label{maxth}
\sli_{\th \ge 0}   {m \choose \th} (h+\ga_C)^{m-\th} \;
\max\{1,\th\} {\ga_C}^{\th}
\end{equation}
Using $\max\{1,\th\} \le 2^\th$, we can bound this sum by $(h+ 3
\ga_C)^m$. Thus the sum over the $m_q$ and $\barm_q$ gives
\begin{equation}\label{guat}
\norm{W_p(V)}_h \le (p-2)! \; {\om_C}^{p-1} \; {\norm{V}_{h'}}^p
\end{equation}
with $h' = h+ 3 \ga_C$. The $1/p!$ in the denominator in
\Ref{Wpexp} cancels the $(p-2)!$, and \Ref{orderP} follows by
summation over $p \ge P+1 \ge 2$. Similarly, to prove \Ref{logbo},
we use \Ref{guat}, bound $(p-2)!/p! \le 1/p$ for $p \ge 2$ and
note that $\norm{W_1(V)}_h = \norm{\mu_C * V}_h$, so that the
bound for the term $p=1$ follows by \Ref{convonorm} and
monotonicity of $\norm{\;\cdot\;}_h$ in $h$. Thus, summing over $p
\ge 1$ gives
\begin{equation}
\norm{W(V)}_h \le \sli_{p \ge 1} \frac{1}{p} \; {\om_C}^{p-1} \,
{\norm{V}_{h'}}^p = - \frac{1}{\om_C} \log \left(1 - (\om_C
\norm{V}_{h'})^p\right).
\end{equation}
Finally, we prove that $h'$ can be replaced by $h''=h+ 2 \ga_C$ if
$\om_C$ is replaced by $2\om_C$ in the bounds in Theorem
\ref{th1}. In the sum over incidence numbers $d_1, \ldots, d_p$,
there is the constraint $d_1+ \ldots +d_p = 2(p-1)$. We thus write
\begin{equation}
\frac{1}{(d_1-1)! \ldots (d_p-1)!}
=
\frac{d_1 \ldots d_p}{d_1! \ldots d_p!}
\end{equation}
and use the arithmetic--geometric inequality, to get
\begin{equation}
d_1 \ldots d_p \le \left( \sfrac{d_1+\ldots +d_p}{p} \right)^p =
2^p \left(1-\sfrac{1}{p}\right)^p \le  2^p \frac{1}{\E} \le
2^{p-1}.
\end{equation}
Then the factors $\max\{1,\th\}$ drop out of \Ref{maxth}, so the
sums over $\th_q$ and $\bar\th_q$ give $2^{m_q+\barm_q}$ instead
of $3^{m_q+\barm_q}$, and hence
\begin{equation}
\norm{W_p(V)}_h \le (p-2)! \; (2\om_C)^{p-1} \;
{\norm{V}_{h+2\ga_C}}^p.
\end{equation}

\subsection{Exponential decay and cumulant bounds}\label{cumusect}
Let $\pseudo (X,X')$ be a pseudometric on $\Xspace$ (i.e.\ satisfy
all properties of a metric except possibly that $\pseudo (X,X') =
0$ implies $X=X'$). A typical example of this situation is if
$\Xspace= \cM \times A$ with $\cM$ a metric space, such as a torus
in real space and $A$ a finite set (such as colour and spin
indices).
\begin{theorem}\label{th2}
Assume that $C$ satisfies \Ref{ipro} and that there are constants
$\tomC$ and $\ellC$ such that for all $X$, $X' \in \Xspace$
\begin{equation}\label{propdecay}
\AbsC (X,X') \le {\ga_C}^2 \, \tomC \; \E^{-\pseudo(X,X')/\ellC}.
\end{equation}
Let $p \ge 2$, $m_1, \ldots , m_p \ge 0$, $\barm_1, \ldots,
\barm_p \ge 0$ such that $m_q+\barm_q >0$ is even for all $q \in
\bN_p$, let $\ul{X}_q = (X_{q,1}, \ldots, X_{q,m_q})$ and
$\ul{Y}_q = (Y_{q,1}, \ldots, Y_{q,\barm_q})$, and let
\begin{eqnarray}
\cG \left((\ul{X}_q,\ul{Y}_q)_{q \in \bN_p}\right) &=&
\truncated{\left\langle \psq^{\barm_1}(\ul{Y}_1) \ps^{m_1}
(\ul{X}_1) ; \ldots ; \psq^{\barm_p}(\ul{Y}_p) \ps^{m_p}
(\ul{X}_p) \right\rangle} \nonu\\ &=& \sfrac{\del^p}{\del \la_1
\ldots \del\la_p} \left.\left( \mu_C * \E^{\sum_q\la_q
\psq^{\barm_q}(\ul{Y}_q) \ps^{m_q} (\ul{X}_q)}
\right)\right\vert_{\la_q=0 \forall q \atop \ps=\psq=0}.
\end{eqnarray}
Then
\begin{equation}\label{2cum}
\cG \left((\ul{X}_q,\ul{Y}_q)_{q \in \bN_p}\right)
\le
(p-2)! \, \tomC^{p-1} \, (3 \ga_C)^{\barm + m} \, \E^{-{1 \over
\ellC} \cL((\ul{X}_q,\ul{Y}_q)_{q \in \bN_p})}
\end{equation}
and
\begin{equation}\label{3cum}
\cG \left((\ul{X}_q,\ul{Y}_q)_{q \in \bN_p}\right)
\le
(p-2)!\,  (2\tomC)^{p-1} \, (2 \ga_C)^{\barm + m}\, \E^{-{1 \over
\ellC} \cL((\ul{X}_q,\ul{Y}_q)_{q \in \bN_p})}
\end{equation}
with $\cL((\ul{X}_q,\ul{Y}_q)_{q \in \bN_p})$ defined as the
minimum of
\begin{equation}
\min\limits_{T \in \cT(\bN_p)} \min \left\{ \sli_{\{q,q'\} \in T}
\pseudo (X_{q,i},Y_{q',j}) : i \in \bN_{m_q}, j \in
\bN_{\barm_{q'}}\right\}
\end{equation}
and
\begin{equation}
\min\limits_{T \in \cT(\bN_p)} \min \left\{ \sli_{\{q,q'\} \in T}
\pseudo (Y_{q,j},X_{q',i}) : j \in \bN_{\barm_{q}}, i \in
\bN_{m_{q'}}\right\}.
\end{equation}
\end{theorem}

\begin{proof}
If we write the monomials as
\begin{eqnarray}
\ps(X_1) \ldots \ps(X_m) &=& \int \dd^m\ul{X'} \pli_{k=1}^m
\de(X_k,X'_k) \ps^m(\ul{X'}) \nonu\\ &=& \int \dd^m\ul{X'} \bA_m
\left(\pli_{k=1}^m \de(X_k,X'_k)\right) \ps^m(\ul{X'})
\end{eqnarray}
with the antisymmetrization operator
\begin{equation}
\bA_m F (X_1, \ldots , X_m) = 
\frac{1}{m!}
\sli_{\pi \in \cS_m} \veps(\pi)
F(X_{\pi(1)}, \ldots, X_{\pi(m)}),
\end{equation}
$\cG$ becomes the field--independent term of the special case of
$\truncated{\langle V;\ldots;V \rangle}$ where $V_q$ consists only
of the term $(\barm_q,m_q)$, with coefficient function
\begin{equation}\label{sapperment}
v^{(q)}_{\barm_q,m_q} (\ul{Y}'_q,\ul{X}'_q)
=
\bA_{m_q} \left[ \pli_{k=1}^{m_q} \de(X_{q,k},X'_k)\right] \;
\bA_{\barm_q} \left[ \pli_{l=1}^{\barm_q} \de(Y_{q,l},Y'_l)\right]
.
\end{equation}
Integrating over the $X'$ and $Y'$ variables only removes the
delta functions; in particular
\begin{equation}\label{intle1}
\sup\limits_{\ul{X}_q,\ul{Y}_q} \int \pli_{q=1}^p \dd^{\barm_q}
\ul{Y}_q' \, \dd^{m_q} \ul{X}_q' \; \abs{v^{(q)}_{\barm_q,m_q}
(\ul{Y}_q',\ul{X}_q')} \le 1
\end{equation}
We now consider the contribution $\cA_T$ of one tree $T \in
\cT(\bN_p)$ in the sum in Theorem \ref{BBFth} to the absolute
value of the connected correlation $\cG$. The only differences to
\Ref{lengthy} are that

\begin{itemize}

\item
there is no sum over $m_q$ and $\barm_q$.

\item
because we now consider the field--independent part
($\ps=\psq=0$), all fields are integrated over; this picks out the
term $A=D$ and $\bar A=\bar D$ in Lemma \ref{lapdet}.

\end{itemize}
The second condition implies that $\cG$ vanishes unless
$m=m_1+\ldots+m_p = \barm_1+\ldots+\barm_p = \barm$, which we
assume from now on. Then
\begin{eqnarray}
\cA_T &\le& \int \dd {\bf X}\, \dd {\bf Y}\; \pli_{\{q,q'\} \in T}
\AbsC(X_q,\bar X_{q'}) \nonu\\ && \pli_{q=1}^p
\abs{v^{(q)}_{\barm_q,m_q} ({\bar X}^{(q)},{\bar
Y}^{(q)},{X}^{(q)},{ X}^{(q)} )}
\\
&& \pli_{q=1}^p \left[ \tst{{m_q \choose \th_q}} \th_q! \;
\tst{{\barm_q \choose \bar\th_q}} \bar\th_q! \;
{\ga_C}^{\barm_q-\bar \th_q + m_q-\th_q} \right]. \nonu
\end{eqnarray}
Here we denoted thoses integration variables on which $\AbsC$
factors depend by $X$, the others by $Y$. By definition, the lines
in the tree can only connect distinct $q$ and $q'$. By
\Ref{propdecay},
\begin{equation}
\pli_{\{q,q'\} \in T} \AbsC(X_q,\bar X_{q'})
 \le
{\ga_C}^{2(p-1)} \tomC^{p-1} \E^{-{1\over\ellC} \sum_{\{q,q'\} \in
T} \pseudo(X_q,X_{q'})}
\end{equation}
By \Ref{inzi}, the factor ${\ga_C}^{2(p-1)}$ combines with the
other powers of $\ga_C$ to ${\ga_C}^{\barm+m}$. By
\Ref{sapperment}, the pseudodistances appearing in the sum are all
of the form $\pseudo (X_{q,i},Y_{q',i'})$, so
\begin{equation}
\sum_{\{q,q'\} \in T} \pseudo(X_q,X_{q'}) \ge \cL
\left((\ul{X}_q,\ul{Y}_q)_{q \in \bN_p}\right).
\end{equation}
We can now bound the integral by $1$ using \Ref{intle1} and then
sum over all trees. Again, the only dependence on the tree left is
in the incidence numbers. As in the proof of Theorem \ref{th1}, we
get factors $2^{m + \barm} \, 2^{p-1}$ or $2^{m + \barm}$,
depending on how we do the bounds. Thus \Ref{2cum} and \Ref{3cum}
hold.

\end{proof}

\noindent In \cite{Salmhofer/Wieczerkowski:1999}, we use these
bounds to derive estimates for norms $\norm{\; \cdot\;}_{h,\ell}$
which keep track of the decay length $\ell$, and also construct a
superrenormalizable version of the Gross--Neveu model using these
norms.

\section{Discussion}
We have seen that the heuristic principle of resumming a graphical
expansion in terms of trees can be realized in different ways, and
only one of them was suited for using Gram estimates. This
nonuniqueness of the tree representation is not surprising because
there is no canonical way of associating a tree with a given
graph. To get the decomposition in Lemma \ref{lpen}, we had to introduce
a particular ordering on the set of all lines to obtain a
well--defined map $G \mapsto T = \Phi(G)$. The BBF interpolation
expansion does not group graphs into disjoint sets associated to
different trees. Instead, the parameters used for the decoupling
of vertices provide tree--dependent weight factors for the graphs. In the
representation \Ref{Schmarrn}, the interpolation parameters are
associated to the lines of the tree and not to the vertices; in
fact, in that approach, interpolation parameters can be avoided
altogether by expanding down the Laplacian in
$\E^{\De_{qq'}+\De_{q'q}}-1$ piece by piece (using that the
Laplacian $\De_{(q,X),q'} = {\delta \over \delta \ps_q (X)} \int
\dd X' \, C(X,X') {\delta \over \delta \psq_{q'} (X')}$ is
nilpotent). Thus, although at the moment not sufficient for
proving convergence, the representation \Ref{nicetry}
may be a good way of
organizing perturbation expansions in practical calculations
because no interpolation integrals are needed.
It is better to have the sign cancellations occur in a determinant 
than to have at the very end a difference of two large numbers 
which are almost equal.

We now discuss (our understanding of) the relation of our approach
to others that have appeared recently.

The construction \cite{Disertori/Rivasseau:1998} 
of the Gross--Neveu model and the many--fermion system is, 
at least technically, rather different from the approach taken here, 
in that it relies on forest formulas that are more explicit
and that seem more closely tied to
the Feynman graph expansion than our tree representations.
Positivity is also used in the technical parts of the proofs in 
\cite{Disertori/Rivasseau:1998}.

The ring expansion invented in
\cite{Feldman/Knoerrer/Trubowitz:1998} is as simple as our
approach as regards the combinatorial and technical complications
in the proof. Very roughly speaking, the operator $R$ introduced
there adds layers to the Feynman graphs, and thus to the spanning
trees, and the condition that $\norm{R} <1$ corresponds to our
condition that $\om_C \norm{V}_h < 1$. A technical difference is
that the expansion in \cite{Feldman/Knoerrer/Trubowitz:1998} is
for the {\em externally connected} functions whereas our proof
deals directly with the connected correlation functions
themselves.

There are some more essential differences. 
Firstly, an advantage of the 
representation in \cite{Feldman/Knoerrer/Trubowitz:1998}
is that $\norm{R} <1$ is, while sufficient, not necessary for
the representation to be defined because the formulas involve
$(1-R)^{-1}$, whose existence only requires that 1 is not an eigenvalue 
of $R$. 

Secondly, Wick ordering
is used in \cite{Feldman/Knoerrer/Trubowitz:1998} to organize the
ring expansion, but it plays no role in our approach. Wick
ordering would simply correspond to dropping the diagonal terms
$\sum_q \De_{qq}$ from our Laplacians. This destroys the
positivity of the matrix $M$ even in the case of the BBF
decoupling. However, the positivity can easily restored by adding
and subtracting the diagonal term and applying the two Laplacians
one after the other, in the same way as we did it in Appendix
\ref{netpos}. This merely changes the Gram constant by a factor 2.
Thus in our approach, Wick ordering could also be used, but it
makes the constants worse.

Thirdly, an advantage of our norm bounds over those in 
\cite{Feldman/Knoerrer/Trubowitz:1998} is that they
are also sharp in the limit $C \to 0$,
where $W(V) \to V$, and $\ga_C \to 0$. Our shifted norm parameter
$h'$ satisfies $h'= h+3 \ga_C \to h$, so that in the limit of no
integration ($C \to 0$), we do not lose anything in the $h$--behaviour. 
In \cite{Feldman/Knoerrer/Trubowitz:1998}, the norm parameter
shifts to $h+1$. % because of binomial factors. 
%
%The bounds in \cite{Feldman/Knoerrer/Trubowitz:1998} are
%stated in terms of a scale parameter $L$, and some constants
%in these bounds diverge in the limit $L \to 1$, which here corresponds
%to the limit $C \to 0$. 

Because our bounds are suitable for $C \to 0$, they stay useful
for $C \sim \dot C \, \Delta t$ even in the limit $\Delta t \to 0$,
and they imply that the renormalization group differential
equation (RGDE)
\begin{equation}\label{rgde}
\dot W = \De_{\dot C} W + \frac12 ({\delta W\over\delta\ps}, \dot
C {\delta W\over\delta \psq})
\end{equation}
is well--defined and has a solution in a ball where $\norm{W}_h$
is small enough, uniformly in $\abs{\Xspace}$. This follows simply
because, by definition, the effective action $W(V)$ is the
solution of the RGDE \Ref{rgde} with the initial condition that 
$W$ equals $V$ at flow time $t=0$.
Of course, we have not used any differential equation techniques
to prove this. 
In particular, our proof does not constitute a nonperturbative
version of Polchinski's method \cite{Po} of proving perturbative renormalizability by integrating differential inequalities.

A Polchinski--type proof of norm bounds similar to ours would
probably give the simplest and most elegant tool in fermionic
constructive field theory. Unfortunately, the proof
in \cite{Brydges/Wright:1988}, which uses differential
inequalities, contains a gap.
This is one of the reasons why we used a discrete technique in this paper, 
to prove a norm bound similar to the one in \cite{Brydges/Wright:1988}.
Our bound is slightly weaker: in \cite{Brydges/Wright:1988}, 
a bound for $\norm{W(V)}_h$ in terms of $ \norm{V}_{h+\ga_C}$, 
i.e., without a factor in front of the $\ga_C$, was stated. 
We believe that the question if and how the gap in the proof in
\cite{Brydges/Wright:1988} can be fixed by a differential equation
argument is related to what the optimal prefactor is. 
This is also why we discussed this prefactor in and
after Theorem \ref{th1}.

One appealing feature of our norm bounds is that every order $p$ in the
expansion of $W$ in terms of $V$ is bounded separately. This makes
it convenient for calculating $W(V)$ to low orders in $V$ and
taking norm bounds of the remainders.

The bounds given here have natural applications in RG studies of
the Gross--Neveu model \cite{Salmhofer/Wieczerkowski:1999} and the
many--fermion problem.

\newpage

\begin{appendix}
\section{The decoupling expansion}\label{BBFapp}
For $\emptyset \ne A \subset \bN_p$, let
\begin{equation}
\tilde \De_{A,q} [M] = \sli_{q'\in A} M_{q'q} (\De_{q'q}+\De_{qq'}
).
\end{equation}
Then, if $M=M^T$,
\begin{equation}
\De_Q [M^{(A,s)}] = \De_A[M] + \De_{Q \setminus A} [M] + s \sli_{q
\in Q \setminus A} \tilde \De_{A,q}[M].
\end{equation}
In particular,
\begin{eqnarray}
\De_Q[M^{(A,1)}] &=& \De_Q[M], \nonu\\ \De_Q[M^{(A,0)}] &=&
\De_A[M]+ \De_{Q \setminus A} [M],
\end{eqnarray}
and for all $ s \in [0,1]$ and all $B$ that satisfy either $B \cap
A = 0$ or $B \subset A$,
\begin{equation}\label{nixs}
\De_B[M^{(A,s)}] = \De_B[M]
\end{equation}
because the constraint $q,q'\in B$ in the definition of $\De_B$
makes off--diagonal terms of type $q \in A$, $q' \not\in A$
impossible. Taylor expansion now gives
\begin{equation} \label{interpolo}
\E^{\De_Q[M]} = \E^{\De_A[M]}\E^{\De_{Q \setminus A} [M]} +
\sli_{q \in Q \setminus A} \tilde \De_{A,q} [M] \ili_0^1 \dd s \;
\E^{\De_Q [M^{(A,s)}]}.
\end{equation}

\begin{lemma}
Let $Q \subset \bN_p$, $M = M^T \in M_p(\bR)$. For $r \ge 1$ let
\begin{equation}
\cS_r(Q) = \{ \qq = (q_1,\ldots,q_r) : q_1 = \min Q, \forall i:
q_i \in Q, q_i \ne q_j \mbox{ if } i \ne j \}.
\end{equation}
Then for all $R \ge 1$,
\begin{eqnarray}\label{ih1}
\E^{\De_Q[M]} &=& \sli_{r=1}^R \sli_{\qq \in \cS_r(Q)} \E^{\De_{Q
\setminus A_r} [M]} \int \pli_{w=1}^{r-1} \dd s_w \tilde
\De_{A_w,q_{w+1}}[M_w]\; \E^{\De_{A_r}[M_r]} \nonu\\ &+& \cR_{R+1}
\end{eqnarray}
with $A_w=\{q_1,\ldots,q_w\}$ and the $M_r$ defined recursively as
$M_1=M$, $M_{w+1} = {M_w}^{(A_w,s_w)}$, and a remainder term
\begin{equation}\label{ih2}
\cR_{R+1} = \sli_{\qq \in \cS_{R+1} (Q)} \int \pli_{w=1}^{R} \dd
s_w \tilde \De_{A_w,q_{w+1}}[M_w]\; \E^{\De_Q[M_{R+1}]} .
\end{equation}
For all $w \in \nato{R}$ and all $B \subset Q \setminus A_w$,
\begin{equation}\label{ih3}
\De_{B} [M_w] = \De_{B} [M],
\end{equation}
and if $M \ge 0$, then $M_w \ge 0$ for all $w \in \nato{R+1}$.
\end{lemma}

\begin{proof}
Induction on $R$, with \Ref{ih1}, \Ref{ih2}, \Ref{ih3}, and $M_w
\ge 0$ for all $w \le R+1$, as the inductive hypotheses. The
statement for $R=1$ is \Ref{interpolo}, with $A=\{q_1\}$. $R
\mapsto R+1:$ In the remainder term, the sum over $\qq \in
\cS_{R+1}$ includes a sum over $q_{R+1} \not\in A_R$. Let $A_{R+1}
= A_R \cup \{ q_{R+1}\} $, and $M_{R+2} =
(M_{R+1})^{(A_{R+1},s_{R+1})}$. Then $M_{R+2} \ge 0$ by Lemma
\ref{interposo}. Now apply \Ref{interpolo} to $\E^{\De_Q[M_{R+1}]}
$. The second summand in \Ref{interpolo} gives the new remainder
term $\cR_{R+2}$. The first summand in \Ref{interpolo} is
\begin{equation}
\E^{\De_{Q \setminus A_{R+1}}[M_{R+1}]}
\E^{\De_{A_{R+1}}[M_{R+1}]}.
\end{equation}
Because $A_R \subset A_{R+1}$, $B=Q \setminus A_{R+1} \subset Q
\setminus A_{R}$, so $B \cap A_{R} = \emptyset$. Thus by
\Ref{nixs},
\begin{equation}
\De_{Q \setminus A_{R+1}}[M_{R+1}]
=
\De_{B}[M_{R}^{(A_R,s_R)}]
=
\De_{B}[M_R].
\end{equation}
By the inductive hypothesis \Ref{ih3}, $\De_{B}[M_R]=\De_{B} [M]$,
hence does not depend on $\sss$, so its exponential can be taken
out of the integral.
\end{proof}

\noin If $R= |Q|$, $\cS_{R+1} (Q) = \emptyset$, so the remainder
term vanishes, and we get
\begin{equation}
\E^{\De_Q[M]} = \sli_{J \subset Q \atop J \ni \min Q} \E^{\De_{Q
\setminus J}[M]} \cK(J)
\end{equation}
where for $|J|=j$,
\begin{equation}
\cK(J) = \sli_{\qq \in \cS_j(J)} \int \pli_{i=1}^{j-1} \dd s_i
\tilde \De_{\{q_1,\ldots,q_i\},q_{i+1}}[M_i]\; \E^{\De_{J}[M_j]}.
\end{equation}
By Lemma \ref{uniconn}, $\connpart{(\E^{\De})}(Q) = \cK(Q)$.
It remains to bring $\cK(\bN_p)$ to the form stated in Theorem
\ref{BBFth} and to show \Ref{normalized}. The conditions in the
sum over sequences in $\cS_p(\bN_p)$ imply that $\cS_p(\bN_p)$ is
the set of all permutations $i \mapsto q_i = \pi(i)$ with
$\pi(1)=1$. The sum in the definition of $\tilde \De_{A,q_i}$ runs
over $q_{v(i)}$ with $v(i) < i$. Thus
\begin{eqnarray}\label{laestig} \connpart{(\E^{\De[M]})} (\bN_p)
=
\sli_{v:\{2,...,p\} \to \{1,..., p-1\} \atop v(i) < i} \sli_{\pi
\in \cS_p \atop \pi(1)=1}
\ili_{[0,1]^{p-1}} 
\dd\sss\; f(\pi,v,\sss) \nonu\\ \pli_{r=2}^p
(\De_{\pi(v(r)),\pi(r)} + \De_{\pi(r),\pi(v(r))} )
M_{\pi(v(r)),\pi(r)} \; \E^{\De[M_p]}
\end{eqnarray}
where $f(\pi,v,\sss) \ge 0$ is a monomial in $\sss$ arising from
the repeated interpolation. We shall not need an explicit
expression for it (it is given in \cite{Mack/Pordt} and needed for
the explicit Gram representation of \cite{Lesniewski:1987}).

The map $v$ is a special case of a predecessor relation defining a
tree: for every $v$ in the above sum, $T_v = \{ \, \{v(i),i\}: i
\in \{2, \ldots, p\}\, \}$ is a tree on $\bN_p$. The map $v
\mapsto T_v$ is injective, but not surjective because of the
particular ordering induced by $v$ (for instance, the tree
$T=\{\{1,3\},\{2,3\}\}$ is not $T_v$ for any $v$ with $v(i) < i$).
On the other hand, every tree on $\bN_p$ is of the form $T_v^\pi=
\{ \, \{\pi(v(i)),\pi(i)\}: i \in \{2, \ldots, p\}\, \}$ for some
$\pi$ and $v$. The Laplacian in \Ref{laestig} does not depend on
$v$, and the product in \Ref{laestig} runs over lines of
$T_v^\pi$. Thus we can reorganize the sums over $v$ and $\pi$ by a
sum over trees $T$ and a sum over $v$, $\pi$ with the constraint
that $T_v^\pi = T$.
Defining
\begin{equation}
\vphi(T,\pi,\sss) = \sli_{v: T_v^\pi=T} f(\pi,v,\sss) \ge 0,
\end{equation}
and $\Pi(T)$ as the set of permutations $\pi$ for which $\pi(1)=1$
and $T^\pi=T$, we get \Ref{BBFrep}.

The proof of \Ref{normalized} is now as given by Battle and
Federbush \cite{Battle/Federbush}: \Ref{BBFrep} holds for any
family of commuting variables $\De_{qq'}$ and matrices $M_{qq'}$.
Let $T_0$ be a fixed tree, $\veps > 0$, and $M_{qq'} = \veps $ if
$\{q,q'\} \in T_0$, $M_{qq'} = 0$ otherwise (in particular,
$M_{qq} =0$). Set $\De_{qq'} = 1/2$. Then \Ref{BBFrep} implies
that
\begin{equation}
\lli_{\veps \to 0} \veps^{-p+1} \connpart{(\E^{\De[M]})} (\bN_p)
=
\int \dd\sss  \;
\sli_{\pi \in \Pi(T_0)} \vphi(T_0,\pi,\sss).
\end{equation}
On the other hand, in the standard representation of the connected
part by a sum over connected graphs,
\begin{equation}
\connpart{(\E^{\De[M]})} (\bN_p)
=
\pli_{q=1}^p \E^{M_{qq}\De_{qq}} \sli_{G \in
\connpart{\cG}(\bN_p)} \pli_{\{q,q'\} \in G} \left( \E^{M_{qq'}
(\De_{qq'}+\De_{q'q})} -1 \right),
\end{equation}
the above choice for $M_{qq'}$ picks out the contribution from the
tree $T_0$, so
\begin{equation}
\veps^{-p+1} \connpart{(\E^{\De[M]})} (\bN_p)  \to 1
\end{equation}
as $\veps \to 0$.

Thus we get the tree expansion for the connected part of
$\E^{\De}$ stated in Theorem \ref{BBFth}. If the $\ps_q$ and
$\psq_q$ are independent fields, the Laplacian $\De[M(T,\pi,s]$
really depends on the permutation $\pi$. After evaluation at
$\ps_q=\ps$ and $\psq_q=\psq$ for all $q$, the $\pi$ dependence
drops out and one gets back the fermionic analogue of the BBF
representation given in \cite{Brydges:1984}.

\section{Proof of Lemma \ref{jomei}}\label{detstuff}

%\section{Gr\" asslich's Zeugs}\label{detstuff}
%\subsection{Proof of Lemma \ref{jomei}}
Using source fields, we have
\begin{equation}\label{quelleli}
V_q(\psq,\ps) = \left[ V_q \left( - \sfrac{\de}{\de\et_q},
\sfrac{\de}{\de\etq_q}\right) \E^{(\etq_q,\ps)+(\psq,\et_q)}
\right]_{\et_q=\etq_q=0}
\end{equation}
Integration, differentiation and evaluation at zero are all
continuous operations on the finite--dimensional Grassmann
algebra, hence interchangeable. The source term factors are in the
even subalgebra, so no signs arise from commuting. Thus the left
hand side becomes
\begin{eqnarray}\label{lhs1}
&& \pli_{q\in Q} V_q \left( - \sfrac{\de}{\de\et_q},
\sfrac{\de}{\de\etq_q}\right) \int \dd \mu_C(\psq',\ps')
\pli_{q\in Q} \E^{(\etq_q,\ps+\ps')+(\psq+\psq',\et_q)} \nonu\\
&=& \pli_{q\in Q} V_q \left( - \sfrac{\de}{\de\et_q},
\sfrac{\de}{\de\etq_q}\right) \E^{ \sum_{q,q'\in Q} (\etq_q,C
\et_{q'}) + \sum_{q\in Q} [(\etq_q,\ps)+(\psq,\et_q)]}
\end{eqnarray}
evaluated at $\et_q=\etq_q=0$. Again using \Ref{quelleli}, the
right hand side becomes
\begin{equation}\label{rhs1}
\left[ \pli_{q\in Q} V_q \left( - \sfrac{\de}{\de\et_q},
\sfrac{\de}{\de\etq_q}\right) \E^{\De} \; \E^{\sum_{r\in Q}
[(\etq_r,\ps_r)+(\psq_r,\et_r)]} \right]_{\et=\etq=0}
\end{equation}
Because all $\De_{qq'}$ commute with one another, $\E^{\De} =
\prod_{q,q'} \E^{\De_{qq'}}$. Because
\begin{equation}
\De_{qq'} \E^{\sum_{r\in Q} [(\etq_r,\ps_r)+(\psq_r,\et_r)]}
=
(\etq_q, C \et_{q'}) \E^{\sum_{r\in Q}
[(\etq_r,\ps_r)+(\psq_r,\et_r)]},
\end{equation}
applying $\E^{\De}$ gives
\begin{equation}
\E^{\De_{qq'}} \E^{\sum_{r\in Q} [(\etq_r,\ps_r)+(\psq_r,\et_r)]}
=
\E^{(\etq_q, C \et_{q'})} \E^{\sum_{r\in Q}
[(\etq_r,\ps_r)+(\psq_r,\et_r)]},
\end{equation}
and therefore \Ref{rhs1} is equal to
\begin{equation}
\left[ \pli_{q\in Q} V_q \left( - \sfrac{\de}{\de\et_q},
\sfrac{\de}{\de\etq_q}\right) \E^{\sum_{q,q'} (\etq_q, C
\et_{q'})} \E^{\sum_{r\in Q} [(\etq_r,\ps_r)+(\psq_r,\et_r)]}
\right]_{\et=\etq=0}.
\end{equation}
If we set $\ps_q=\ps$ and $\psq_q=\psq$ for all $q$, we get the
last line of \Ref{lhs1}.

\section{The direct resummation}
\subsection{Penrose's proof of Lemma \ref{lpen}}\label{penapp}
Define a map $\Phi: \cG_c(\bN_p) \to \cT(\bN_p)$ as follows. Let
$G\in \cG_c(\bN_p)$. For a vertex $q\in \bN_p \setminus \{1\}$ let
$l_q$ be the length of a shortest path connecting it to the vertex
$1$. This partitions $\bN_p$ into sets $A_l$ of vertices with
distance $l$ to $1$. Delete all lines $\{q,q'\}$ for which $q \in
A_l$ and $q' \in A_l$, for all $l \ge 1$. Call the resulting graph
$G'$. Then every line of $G'$ goes from $A_k$ to $A_{k+1}$ for
some $k$. Also, $G'$ is still connected, thus for every vertex $q
\ge 2$, the set of lines reaching $q$, $R_q=\{ \{q,q'\} \in G'\}$,
is nonempty. Delete all lines of $R_q$ except the one with the
smallest $q'$ from $G'$. The resulting graph is connected and has
$q-1$ lines. Thus it is a tree $T$. Let $\Phi(G)=T$.

For a tree $T$, $\Phi(T)=T$, so the map $\Phi$ is surjective. The
decomposition given in Lemma \ref{lpen} is the decomposition into
preimages
\begin{equation}
\cG_c(\bN_p) = \bigcup_{T\in \cT(\bN_p)} \Phi^{-1} (\{T\}).
\end{equation}
To get $\Phi^{-1} (\{T\})=\{ G \in \cG_c(\bN_p) : \Phi(G)=T\}$,
one only has to reverse the above algorithm: let $T$ be any tree.
Group the vertices into sets $A_l$ of distance $l$ from $1$ (i.e.\
root the tree at $1$). Let $H^*(T)$ be the graph containing all
the following lines: for $q \ge 2$ let $\th$ be the unique line of
$T$ connecting $q$ to a lower vertex $q'$; all lines $\{q'',q\}$
with $q'' > q'$ belong to $H^*(T)$. For $l \ge 1$, all lines
$q,q'$ with $q \in A_l$ and $q'\in A_l$ belong to  $H^*(T)$.

By construction, all subsets $H$ of $H^*(T)$ satisfy $\Phi(T\cup
H) = T$, and if $G$ is a connected graph containing any line not
in $H^*(T)$, then $\Phi(G) \ne T$.

\subsection{The matrix structure} \label{netpos}
In this section, we show that the matrices $M$ associated to the
direct resummation are band matrices and then provide examples
where they have negative eigenvalues.

\begin{figure}
\epsfxsize=3in \centerline{\epsffile{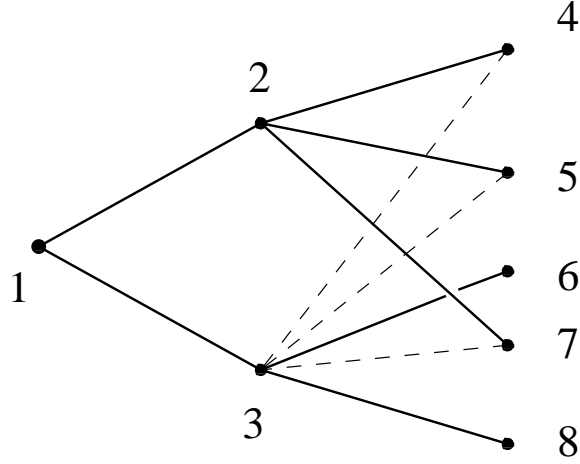}} \caption{A tree
$T$ on $\bN_8$ (solid lines). The dashed lines are those lines of
$H^*(T)$ that connect $V_1=\{2,3\}$ to $V_2=\{4,5,6,7,8\}$. }
\label{treefig}
\end{figure}

We first introduce a natural ordering on the vertex set $\bN_p$.
Let $V_0=\{1\}$, and for $ k \ge 1$ let $V_k $ be the set of
vertices with distance $k$ from $1$ (measured in steps when going
over tree lines). $V_1$ is ordered by the usual ordering on $\bN$.
The set $V_2$ is ordered as follows: First, take the vertices $q$
with $\pi(q)$ the smallest element of $V_1$, and order them in a
similar way as we ordered $V_1$, etc. In the example shown in
Figure \ref{treefig}, this means that the ordering of $V_2$ is
$(4,5,7;6,8)$.

Recall that the lines of $H^*(T)$ are all those that connect
vertices $q,q'$ with $q \in V_k$ and $q' \in V_k$, and those that
connect $V_k$ and $V_{k+1}$ and that are compatible with the
minimality of $T$. The point of the above ordering is that the
second condition is simple: for instance, in the example in Figure
\ref{treefig}, the only allowed lines between $V_1$ and $V_2$ are
those connecting the set $\{4,5,7\}$ to $3$. Connecting $6$ (or
$8$) to $2$ is not allowed by construction of $H^*(T)$.

In the ordering on the vertices just introduced, the matrix
$M^{(T)}$ thus takes the block form (labelled by $1$, $2$, $3$,
and the sets $W_1=\{4,5,7\}$ and $W_2=\{6,8\}$)
\begin{equation}\label{weia}
\left(
\begin{array}{lllll}  
1 & s_{12} & s_{13} & 0 & 0 \\ s_{12} & 1 & 1 & \si_1 & 0 \\
s_{13} & 1 & 1 & 1 & \si_2 \\ 0 & \si_1^T & 1 & 1 & 1 \\ 0 & 0 &
\si_2^T & 1 & 1
\end{array}
\right)
\end{equation}
The blocks denoted by $1$ are matrices with all entries equal to one.
There are $1$'s in the diagonal because all lines with $q=q'$
appear and because all lines from $V_k$ to $V_k$ appear. The
matrices $\si_i$ comprise the $s$--factors from the second layer
of the tree.

For general trees, the matrix $M^{(T)}$ is a block matrix of band
form because the only lines allowed in $H^*(T)$ go either from
$V_k$ to $V_k$ (diagonal blocks) or from $V_k$ to $V_{k+1}$
(blocks adjacent to the diagonal).

Such matrices are typically not positive: already for $p=3$ and
the tree $T=\{ \{1,2\},\{1,3\}\}$ (which corresponds to the left
upper corner of the matrix in \Ref{weia}) and the particular
values $s_{12}=1$ and $s_{13}=0$,
\begin{equation}\label{nixpos}
\det \left(\matrix{1&1&0\cr 1&1&1 \cr 0&1&1 } \right) = -1.
\end{equation}
In this example, positivity is easy to repair: if all diagonal
elements are replaced by $2$, the matrix is just minus the
one--dimensional discrete Laplacian, hence positive. Thus the
matrix in \Ref{nixpos} can be written as a difference
\begin{equation}
\left(\matrix{2&1&0\cr 1&2&1 \cr 0&1&2 } \right)
-
\left(\matrix{1&0&0\cr 0&1&0 \cr 0&0&1 } \right)
\end{equation}
of two Gram matrices. The first one has Gram constant $\sqrt{2}$
by Lemma \ref{goodguy} and the second one Gram constant $1$. Upon
iteration of the application of the Laplacian in two steps, the
two Gram constants add up, so effectively, $\ga_C$ is replaced by
$(1+ \sqrt{2})\ga_C$. Similar tricks work for individual trees,
but do not seem to yield bounds that are uniform in $T$.

\section*{Acknowledgements}
We would like to thank David Brydges, Joel Feldman, Horst Kn\"
orrer, Andreas Pordt, and Eugene Trubowitz for discussions, and the
Forschungsinstitut f\" ur Mathematik, ETH Z\" urich, for its support.
M.S.\ would also like to thank Xue--Ping Wang and the colleagues
at the Mathematics Department of the University of Nantes for
their hospitality during a visit where part of this work was done.

\end{appendix}

\bibliographystyle{plain}

\end{document}